\documentstyle[epsfig,amsmath,12pt]{article}
\begin{document}





\pagestyle{empty}

\renewcommand{\thefootnote}{\fnsymbol{footnote}}


\begin{flushright}
{\small
SLAC--PUB--10176\\
September 2003\\}
\end{flushright}

\vspace{.8cm}

\begin{center}
{\large\bf
Hadronic $B$ Decays with {\sc BaBar}\footnote{Work supported by
Department of Energy contract  DE--AC03--76SF00515.}}

\vspace{1cm}

P. Robbe\\
Laboratoire d'Annecy-le-Vieux de Physique des Particules,
74941 Annecy-le-Vieux, France\\

\medskip
for the {\sc BaBar} Collaboration\\
Stanford Linear Accelerator Center, Stanford University, Stanford,
California 94309

\end{center}

\vfill

\begin{center}
{\large\bf
Abstract }
\end{center}

\begin{quote}
We report about the studies of the decay channels $B^-\to D^0 K^-$,
$B^0\to D^{*-} a_1^+$ and $B^0\to D_s^{(*)-} \pi^+$ with a sample of
62 $\times$ $10^6$ $\Upsilon(4S)$ decays into $B$ meson pairs
collected with the {\sc BaBar} detector at the {\sc Pep} II
asymmetric $e^+ e^-$ collider.
\end{quote}

\vfill

\begin{center}
\textit{Contributed to}  \textit{Beauty 2002: $8^{th}$ International
Conference on $B$ physics at Hadron Machines}\\
\textit{Santiago de Compostela, Spain} \\
\textit{June 17 - June 21, 2002}\\
\end{center}


%
%
\newcommand{\x}{\cdot}
\newcommand{\ra}{\rightarrow}
%
%
%

\vspace{4ex}


%
%

%
%
%


%
\setlength{\oddsidemargin}{0 cm}
\setlength{\evensidemargin}{0 cm}
\setlength{\topmargin}{0.5 cm}
\setlength{\textheight}{22 cm}
\setlength{\textwidth}{16 cm}
\setcounter{totalnumber}{20}

\clearpage\mbox{}
\clearpage

\pagestyle{plain}
\setcounter{page}{1}
%
The measurement of the $CP$-violating phase of the Cabibbo-Kobayashi-Maskawa
(CKM) matrix \cite{ref:ckm} is an important part of the present scientific program
in particle physics. Theoretically clean measurements of the angle $\beta$
of the unitarity triangle exist \cite{ref:beta} but there are no such measurements of the
two other angles ($\alpha$ and $\gamma$). The determination of these
two angles would check the validity of the CKM mechanism in the
explanation of the $CP$-violation. Theoretically clean measurements
of $\gamma$ and $\sin \left( 2\beta + \gamma \right)$ can be obtained
from the study of the decay modes $B\to D^0 K$, $B^0\to D^{*-} a_1^+$
and $B^0\to D_s^{(*)+}\pi^-$.
In this paper we present the results of the measurements of the
branching fractions of the decay modes, as a preliminary step towards
the measurements of the $\gamma$ angle.

\section{Detector and data sample}

The data were collected in the years 1999-2001 with the {\sc BaBar}
detector at the {\sc Pep}-II asymmetric $e^+$ (3.1 GeV) - $e^-$ (9 GeV)
storage ring. The {\sc BaBar} detector is a large-acceptance solenoidal
spectrometer (1.5 T) described in detail elsewhere~\cite{ref:babar}.
The analyses described below make use of charged track and
$\pi^0$ reconstruction and charged particle identification.
Charged particle trajectories are measured by a 5-layer double-sided
silicon vertex tracker (SVT) and a 40-layer drift chamber (DCH), which also
provide ionisation measurements ($dE/dx$) used for particle
identification. Photons and electrons are measured in the electromagnetic
calorimeter (EMC), made of 6580 thallium-doped CsI crystals constructed
in a non-projective barrel and forward endcap geometry. Charged $K/\pi$
separation up to 4 GeV/c in momentum is provided by a detector of internally
reflected Cherenkov light (DIRC), consisting of 12 sectors of quartz
bars that carry the Cherenkov light to an expansion volume filled with water
and equipped with 10751 photomultiplier tubes.

\section{$B^-\to D^0 K^-$}

The study of this decay channel can lead to a clean measurement of
the $\gamma$ angle \cite{ref:gamma1,ref:gamma2}. $B^-\to D^0 K^-$ decays are obtained with
a color-allowed, $V_{us}$ suppressed diagram. $B^-\to \bar{D}^0 K^-$ 
decay modes also exist and are due to a color- and $V_{ub}$ suppressed diagram.
If $D^0$ decays into a $CP$ eigenstate such as $D^0_{CP}\to K^-K^+$, the
decay $B^-\to D^0_{CP} K^-$ can be obtained with both processes.
Knowing the decay amplitudes of all three possibilities ($B^-\to D^0 K^-$,
$B^-\to \bar{D}^0 K^-$ and $B^-\to D^0_{CP} K^-$), it is possible
to measure $2\gamma$ and then $\gamma$ up to discrete ambiguities.

The $CP$ asymmetry ${\cal A}_{CP}$:
\begin{equation}
{\cal A}_{CP} = \frac{{\cal B}\left(B^+\to D^0_{CP} K^+ \right)
			-{\cal B}\left(B^-\to D^0_{CP} K^- \right)}
		{{\cal B}\left(B^+\to D^0_{CP} K^+ \right)
			+{\cal B}\left(B^-\to D^0_{CP} K^- \right)}
\end{equation}
is related to the $\gamma$ angle. It is expected to be of the order
of 10 \% in the Standard Model.

$B^-\to D^0 K^-$ decay modes are expected to have a branching fraction
10 times lower than the branching fraction for $B^-\to D^0 \pi^-$
(${\cal B}\left( B^-\to D^0 \pi^- \right) = \left(5.3\pm 0.5 \right)
\times 10 ^{-3}$ \cite{ref:pdg}) which constitutes the main background
source of this analysis. The capability of the DIRC to distinguish
between pions and kaons will then be very important. Moreover
the interesting $D_{CP}^0$ decay modes
are Cabibbo suppressed and have small branching fractions
(${\cal B}\left( D_{CP}^0\to K^+ K^- \right) = \left(4.12\pm 0.14\right)
\times 10^{-3}$ \cite{ref:pdg}). The large data sample available
at {\sc BaBar} will also be useful.

For this analysis, $D^0$ candidates reconstructed in the decay modes
$D^0\to K^- \pi^+$, $D^0\to K^-\pi^+\pi^0$, $D^0\to K^- \pi^+ \pi^-
\pi^+$ and $D_{CP}^0 \to K^- K^+$ are combined with a prompt charged
track $h^-$ which creates Cherenkov light in the DIRC. The
effective mass of the $B^-$ candidate is calculated using the kaon
mass hypothesis so both $B^-\to D^0 K^-$ and $B^-\to D^0 \pi^-$ are
reconstructed.

For each $B$ candidate, two variables are calculated using the fact
that $B$ mesons are produced in pairs, and are almost at rest in the 
$\Upsilon(4S)$ frame:
\begin{alignat}{1}
m_{ES} = & \, \sqrt{\left(\frac{1}{2}\sqrt{s}\right)^2 - 
\vec{p}^{*2}} \\
\Delta E = & \, E^* - \frac{1}{2}\sqrt{s} 
\end{alignat} 
Signal $B^-\to D^0 K^-$ events will accumulate in a region of the
$m_{ES} - \Delta E$ plane which is centered on $\Delta E = 0$ 
and $m_{ES} = 5.28 \ {\rm GeV/c^2}$ (the nominal $B$ mass) whereas background $B^-\to
D^0 \pi^-$ events will accumulate in a region shifted to positive
$\Delta E$ values but at the same $m_{ES}$ values since $m_{ES}$ 
depends only on the laboratory 3-momentum.

The number of signal $B^-\to D^0 K^-$ events is computed with an extended
maximum likelihood fit which makes use of the position of the $B$ candidate
in the $m_{ES} - \Delta E$ plane and of the Cherenkov angle of the prompt
track $h^-$ to distinguish between $B^-\to D^0 K^-$ events, $B^-\to D^0 \pi^-$
events, peaking background and combinatorial background events. 

\begin{figure}[htb]
  \begin{center}
    \scalebox{0.37}{\includegraphics{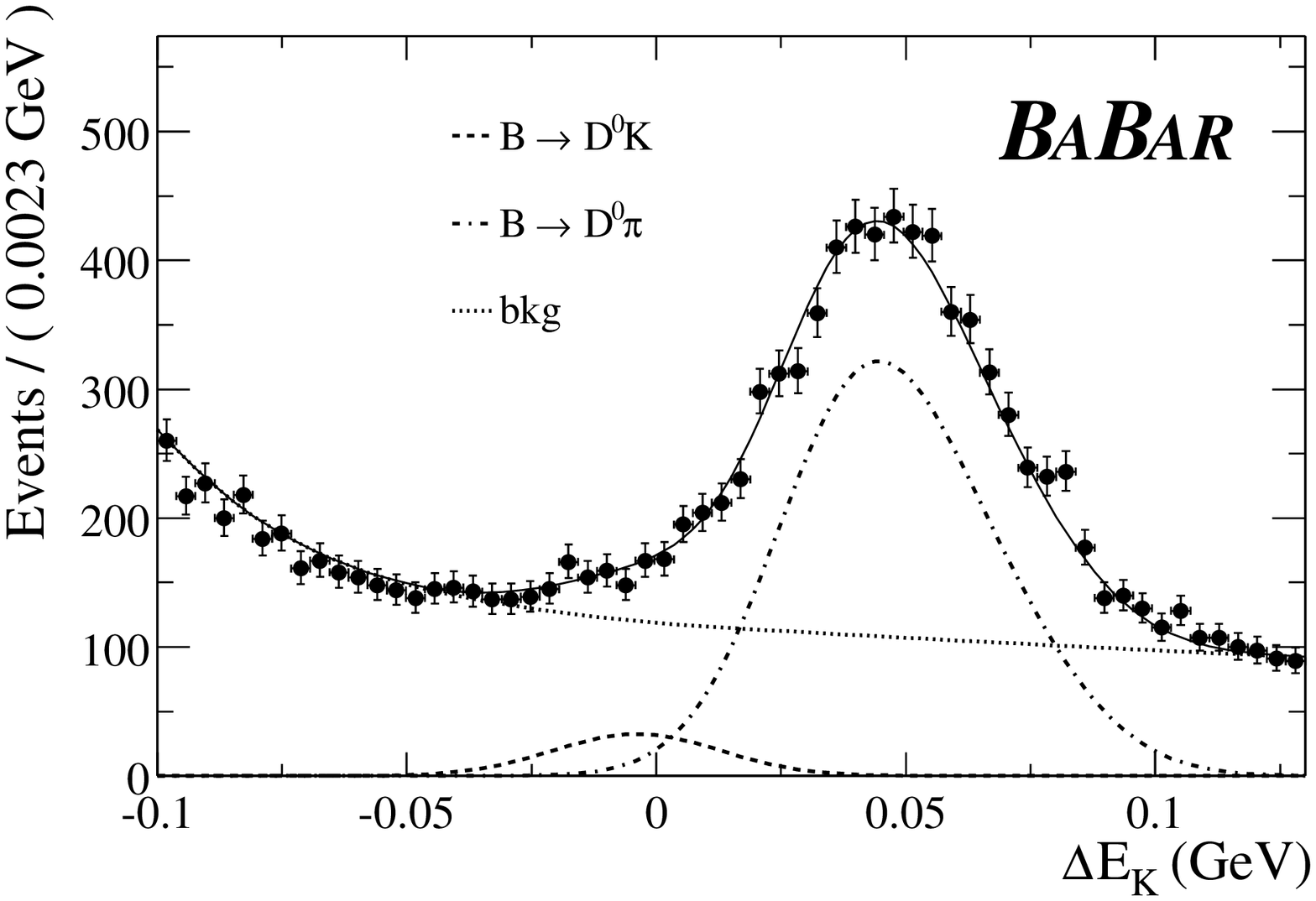}\includegraphics{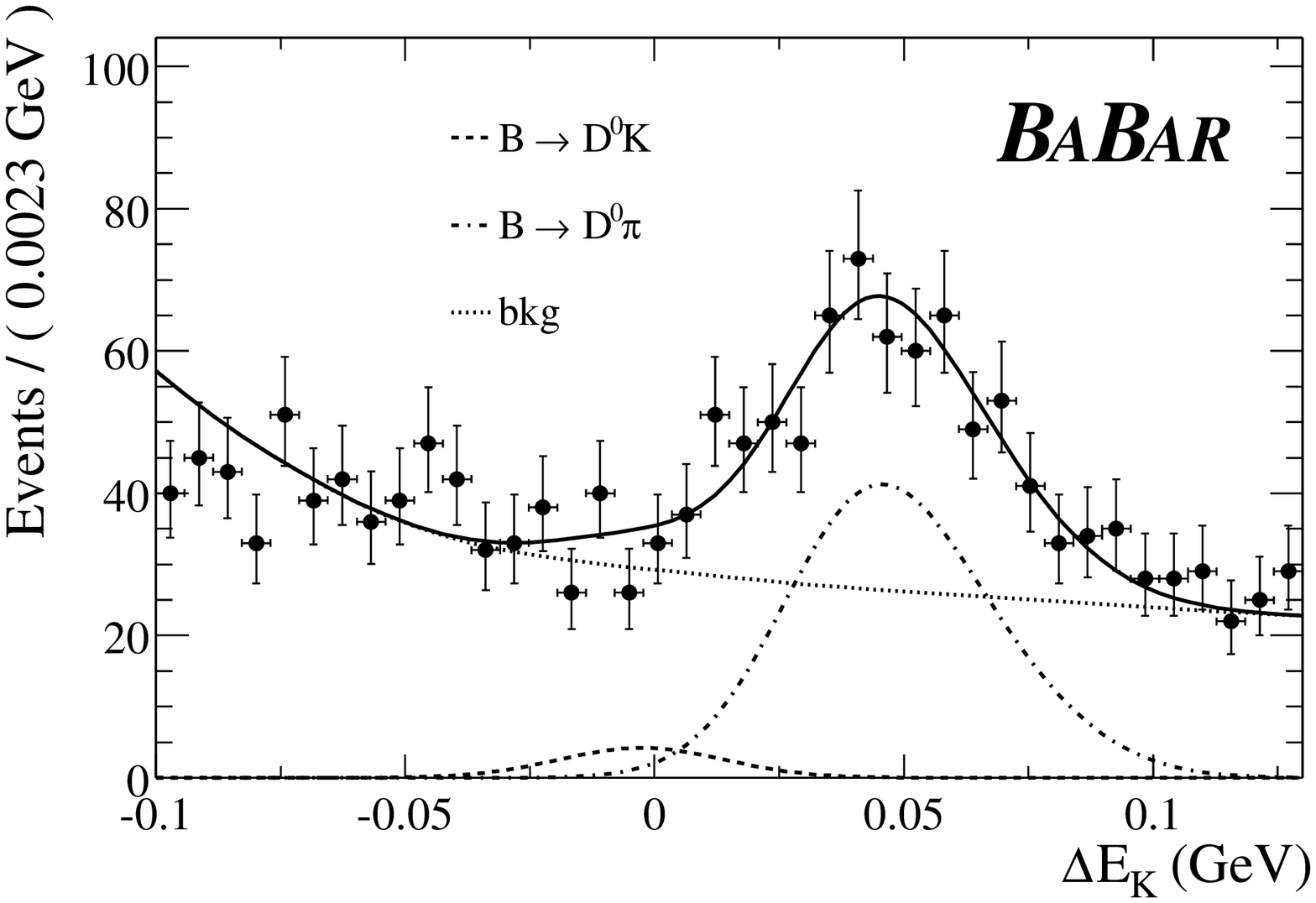}}
    \caption{$\Delta E$ distribution for $B^- \to D^0 h^-$ candidates, with 
$D^0\to K^- \pi^+$ (left) and $D_{CP}^0\to K^- K^+$ (right)}
\label{Fi:dkall}
\end{center}
\end{figure}

Fig. \ref{Fi:dkall} shows the $\Delta E$ projections for all $B^- \to D^0 h^-$
candidates reconstructed in a sample of 56.4 fb$^{-1}$ on-resonance data, with
$D^0\to K^-\pi^+$ (left) and $D_{CP}^0 \to K^- K^+$ (right) decay
modes. On each projection the fitted distribution and the
contributions to the total function of $B^-\to D^0 K^-$, $B^-\to D^0 \pi^-$ and
background events are overlaid.

\begin{figure}[htb]
  \begin{center}
    \scalebox{0.37}{\includegraphics{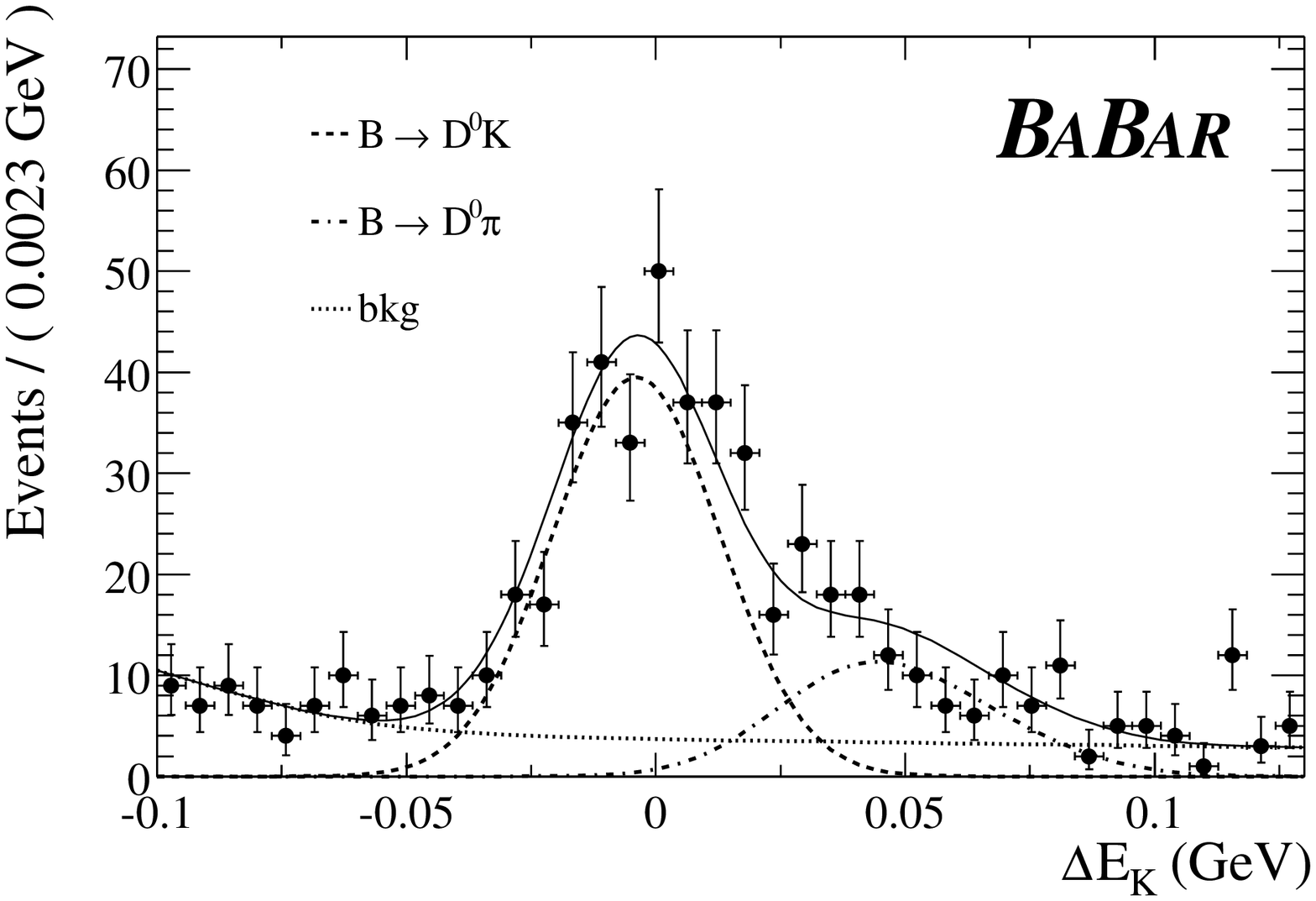}\includegraphics{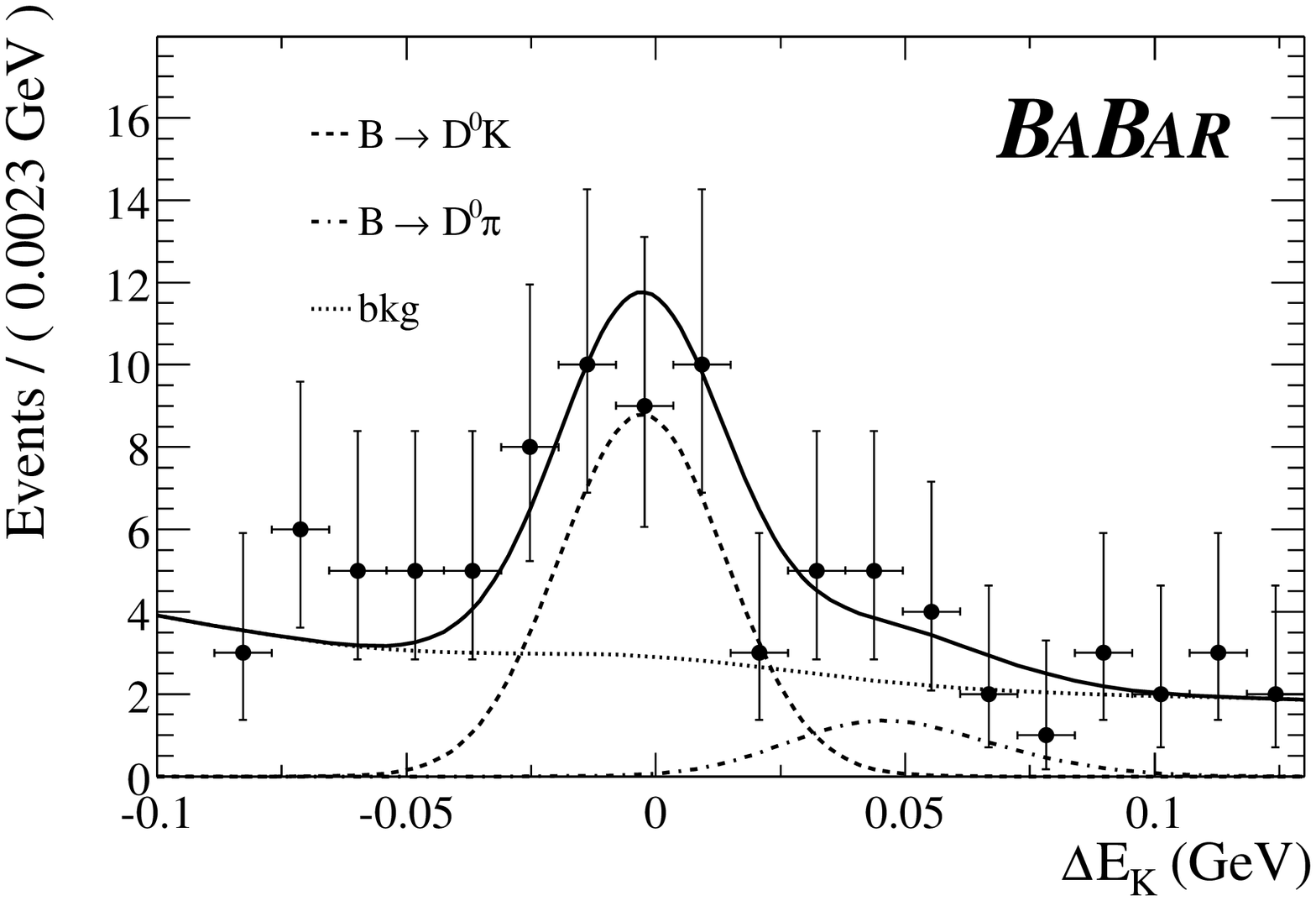}}
    \caption{$\Delta E$ distribution for $B^- \to D^0 h^-$ candidates, with 
$D^0\to K^- \pi^+$ (left) and $D_{CP}^0\to K^- K^+$ (right) with tight kaon identification for the
prompt track $h^-$}
\label{Fi:dkktight}
\end{center}
\end{figure}

A clear evidence of the signal for $B^-\to D^0 K^-$ is obtained requiring tight kaon 
identification criteria on the prompt track $h^-$. The corresponding $\Delta E$
projections are shown on Fig. \ref{Fi:dkktight}.

The branching fractions are extracted from these fits on the same data sample. 
The ratio to the $B^-\to D^0 \pi^-$ branching fraction is measured equal to:
\begin{equation}
R = \frac{{\cal B}\left( B^- \to D^0 K^- \right)}
{{\cal B}\left( B^- \to D^0 \pi^- \right)} = 
\left(8.31 \pm 0.35 ({\rm stat}) \pm 0.13 ({\rm syst})\right) \%
\end{equation}
for non {\em CP} modes and to:
\begin{equation}
R_{CP} = \frac{{\cal B}\left( B^- \to D_{CP}^0 K^- \right)+{\cal B}\left( B^+ \to D_{CP}^0 K^+ \right)}
{{\cal B}\left( B^- \to D_{CP}^0 \pi^- \right)+{\cal B}\left( B^+ \to D_{CP}^0 \pi^+ \right)} = 
\left(8.4 \pm 2.0 ({\rm stat}) \pm 0.8 ({\rm syst})\right) \%
\end{equation}
for {\em CP}-even modes ({\em ie} $D^0\to K^-K^+$).

The $CP$ asymmetry has been found equal to:
\begin{equation}
{\cal A}_{CP} = \frac{{\cal B}\left( B^- \to D_{CP}^0 K^- \right)-{\cal B}\left( B^+ \to D_{CP}^0 K^+ \right)}
{{\cal B}\left( B^- \to D_{CP}^0 K^- \right)+{\cal B}\left( B^+ \to D_{CP}^0 K^+ \right)} = 
0.15 \pm 0.24_{-0.08}^{+0.07}
\end{equation}

\section{$B^0\to D^{*-} a_1^+$}

$B^0$ mesons can decay either into $D^{*-} \pi^+$ (Cabibbo-Allowed diagram)
or into $D^{*+} \pi^-$ (Cabibbo and $V_{ub}$ suppressed diagram). Since $B^0$
mesons can also oscillate to $\bar{B}^0$, the time dependant evolutions
of $B^0\to D^{*-}\pi^+$ and $B^0\to D^{*+}\pi^-$ are related to $\sin\left(2\beta+\gamma
\right)$\cite{ref:babarbook}. This method requires a lot of events to lead to a precise
measurement. It may be interesting to use the similar $B^0\to D^{*-}a_1^+$ 
decay mode which has a larger branching fraction: from \cite{ref:pdg},
${\cal B}\left(B^0\to D^{*-} \pi^+\right) = \left(2.76\pm 0.21\right)\times 10^{-3}$
and 
${\cal B}\left(B^0\to D^{*-} a_1^+ \right)= \left(1.30\pm 0.27\right)\%$.

In order to reconstruct even more events, the analysis described here makes use of 
a partial reconstruction technique \cite{ref:dstara1}, using only the soft pion from the $D^{*-}$ decay
and the $a_1^+$. With respect to the full reconstruction technique, it has thus
no penalty due to the branching fractions of the reconstructed $D^0$ decay modes.
Since the soft pion in the $D^*$ decay has a low momentum, it is very often only
reconstructed in the SVT and the analysis requires a good stand-alone track
reconstruction capability of this device.

In the decay chain $B^0\to D^{*-} a_1^+, D^{*-}\to \bar{D}^0 \pi^-$, only 
the $a_1^+$ and the slow $\pi$ from the $D^*$ decay are reconstructed. 
$a_1^+$ is only reconstructed in the decay mode $a_1^+\to \rho^0 \pi^+$
whose branching fraction is supposed to be equal to 49.2 \%. The remaining 12 
parameters are determined by applying the constraints of 4-momentum conservation to the 
$B$ and $D^*$ decay, the invariant masses of the $B$, $D^*$ and $D^0$
and the $B$ energy in the Center of Momentum frame, that is to say the half
of the Center of Mass energy. By applying the beam energy and $B$ mass constraints, 
the angle between the $B$ and the $a_1$ momentum can be computed. The $B$ 
4-momentum is known up to an azimuthal angle $\phi$ around the $a_1$ momentum.
$\phi$ is the only unknown parameter. The missing mass $m_{miss}$ is computed
averaging over $\phi$. 

For signal events, $m_{miss}$ peaks at the nominal $D^0$ mass but the 
$m_{miss}$ distribution is broader for background events. A large
fraction of background events come from continuum events. This type
of background is rejected in this analysis with a Neural Network
algorithm \cite{ref:salvatore} using the different topologies between continuum events
which have a jet-like structure and $B\bar{B}$ events which are
more spherical. 

\begin{figure}[htb]
  \begin{center}
    \scalebox{0.42}{\includegraphics{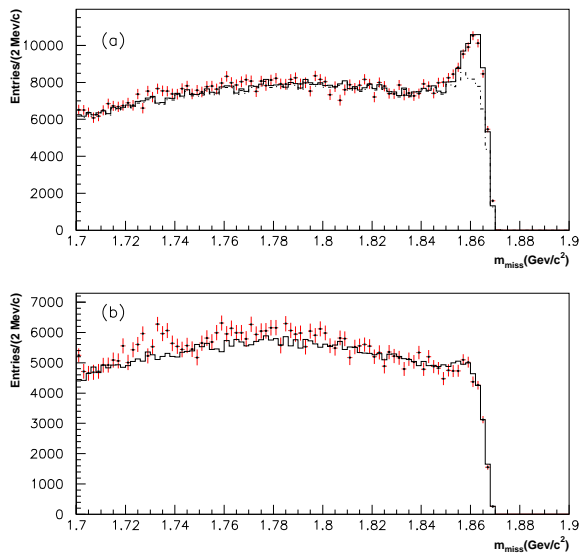}}
    \caption{(a) $m_{miss}$ distribution of continuum-substracted on-resonance data events
(data points), $B\bar{B}$ background MC events (dashed histogram) and $B\bar{B}$ background
plus signal events (solid histogram) for ``right-sign'' $a_1 \pi$ combinations. The
histograms are the result of the fit procedure described in the text. (b) Same distributions
for ``wrong-sign'' $a_1 \pi$ combinations}
\label{Fi:miss}
\end{center}
\end{figure}

Fig. \ref{Fi:miss} shows the missing mass distributions obtained with a sample
of 20.6 ${\rm fb}^{-1}$ on-resonance data and 2.5 ${\rm fb}^{-1}$ off-resonance data,
for opposite sign combinations (top, $a_1^+ - \pi^-$) and same sign combinations (bottom, $a_1^+
- \pi^+$) where no signal is present.
In the distributions presented, the distributions obtained from the off-resonance
data are subtracted from that obtained from data recorded at the $\Upsilon(4S)$
peak. To compute the total number of signal events, the resulting distribution
for opposite sign combinations is fitted with a linear combination of a distribution 
from background $B$ Monte-Carlo events and a distribution from signal Monte-Carlo
events. This procedure yields an estimated signal of $18427\pm 1200$ events. 

The branching ratio resulting from this analysis is found equal to:
\begin{equation}
{\cal B}\left(B^0\to D^{*-} a_1^+\right) = 1.20 \pm 0.07  {\rm (stat)}
\pm 0.14 {\rm (syst)}\ \%
\end{equation}
An additional systematic bias due to the unknown presence
of background $B\to D^{**} a_1$ decays has to be added to the systematic error. 
If $B^{**} = {\cal B}\left(B\to D^{**} a_1\right) \times 
{\cal B}\left(D^{**}\to D^* \pi\right)$ this bias is:
\begin{equation}
\sigma = \left(^{+0}_{-0.05\times B^{**} / 0.35 \% }\right) \ \%
\end{equation}

\section{$B^0\to D_s^{(*)+} \pi^-$}

The determination of $\sin\left(2\beta+\gamma\right)$ with 
$B^0\to D^{*} \pi$ or $B^0\to D^{*} a_1$ mentioned in the previous section
requires the knowledge of the ratio $\lambda$ between the two decay amplitudes of the allowed
and suppressed processes:
\begin{equation}
\lambda = \frac{{\cal A}\left(B^0\to D^{*+}\pi^-\right)}
{{\cal A}\left(B^0\to D^{*-} \pi^+\right)}
\end{equation}

$\lambda$ cannot be measured directly because the two processes cannot be distinguished
experimentally. A way to measure $\lambda$ is to study the decay $B^0\to D_s^{(*)+}\pi^-$
whose branching fraction is related to $\lambda$ \cite{ref:dunietz}:
\begin{equation}
{\cal B}\left(B^0\to D_s^{(*)+}\pi^-\right) =
\frac{{\cal B}\left(B^0\to D^{(*)-}\pi^+\right)}
{\cos\theta_{\rm cabibbo}^2}
\left(\frac{f_{D_s^{(*)}}}{f_{D^{(*)}}}\right)^2
\lambda^2
\end{equation}
where $f_{D_s^{(*)}}$ and $f_{D^{(*)}}$ are the decay constants of
$D_s^{(*)}$ and $D^{(*)}$. The decay $B^0\to D_s^{(*)+} \pi^-$ can also
be used for a measurement of $\left| V_{ub} / V_{cb} \right|$ \cite{ref:kim}.

$B^0\to D_s^{(*)+}\pi^-$ candidates are fully reconstructed, with $D_s$ candidates
reconstructed in the decay modes: $D_s^+\to \phi \pi^+$, $D_s^+ \to K^{*0} K^+$
and $D_s^+\to K_s K^+$ with $\phi \to K^+ K^-$, $K^{*0}\to K^+ \pi^-$ and
$K_s\to \pi^+ \pi^-$. $D_s^*$ candidates are reconstructed in the decay mode
$D_s^{*+}\to D_s^+ \gamma$. Background from continuum events is rejected
using a Fisher discriminant \cite{ref:fisher} and the cosine of the angle between the thrust
axis of the reconstructed $B$ and the thrust axis of the remaining tracks of the event.
For jet-like continuum events, the two thrusts are back to back whereas for
$B\bar{B}$ events, the cosine distribution is flat. The selection algorithm \cite{ref:dspi} makes
use of kaon identification, $D_s^+ \pi^-$ vertex probability and the helicity
angle in the decays $D_s^+\to \phi \pi^+$ and $D_s\to K^{*0} K^+$ in order
to reduce the combinatorial background from continuum and $B\bar{B}$ 
events. 

\begin{figure}[htb]
  \begin{center}
    \scalebox{0.5}{\includegraphics{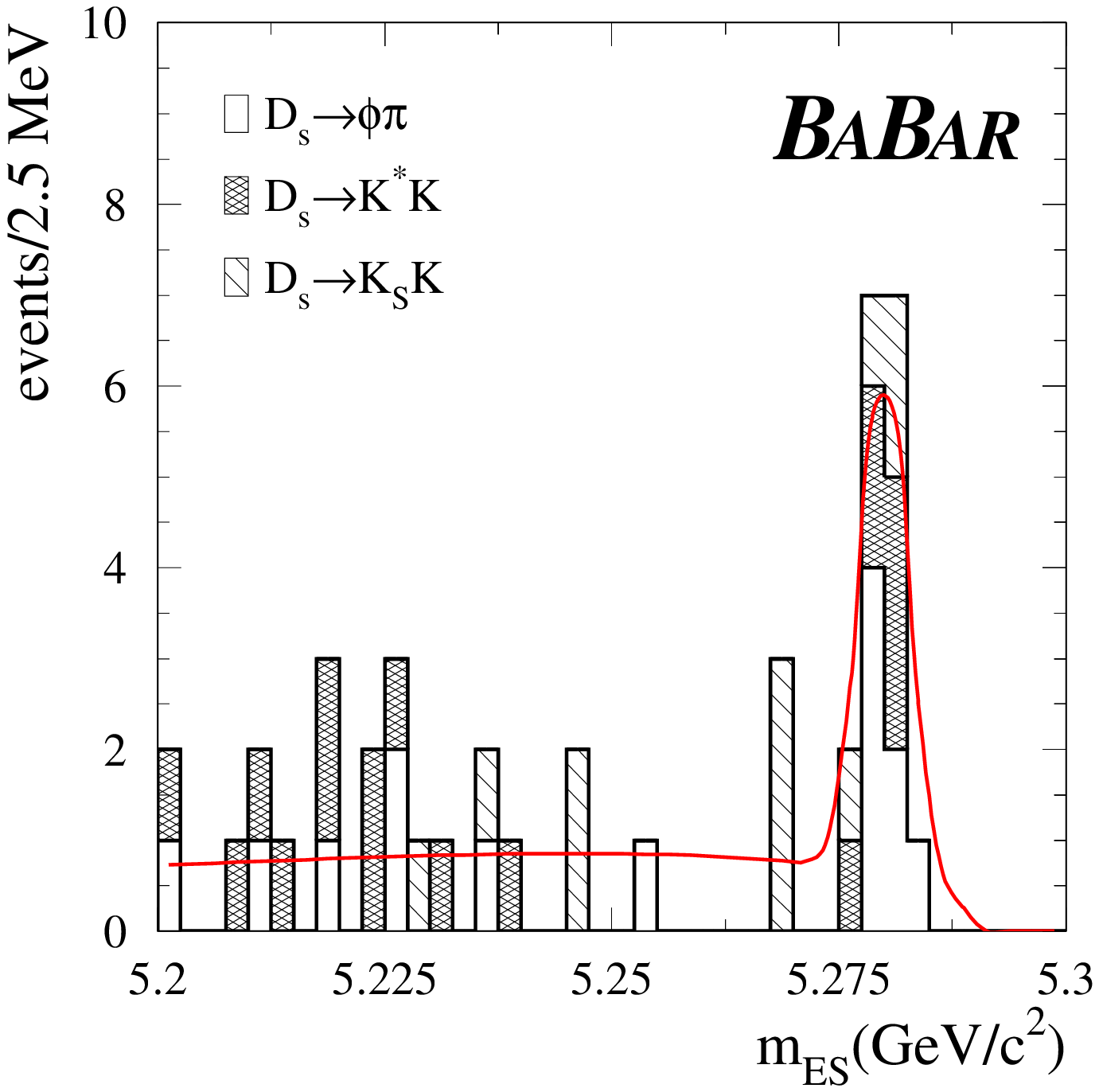}\includegraphics{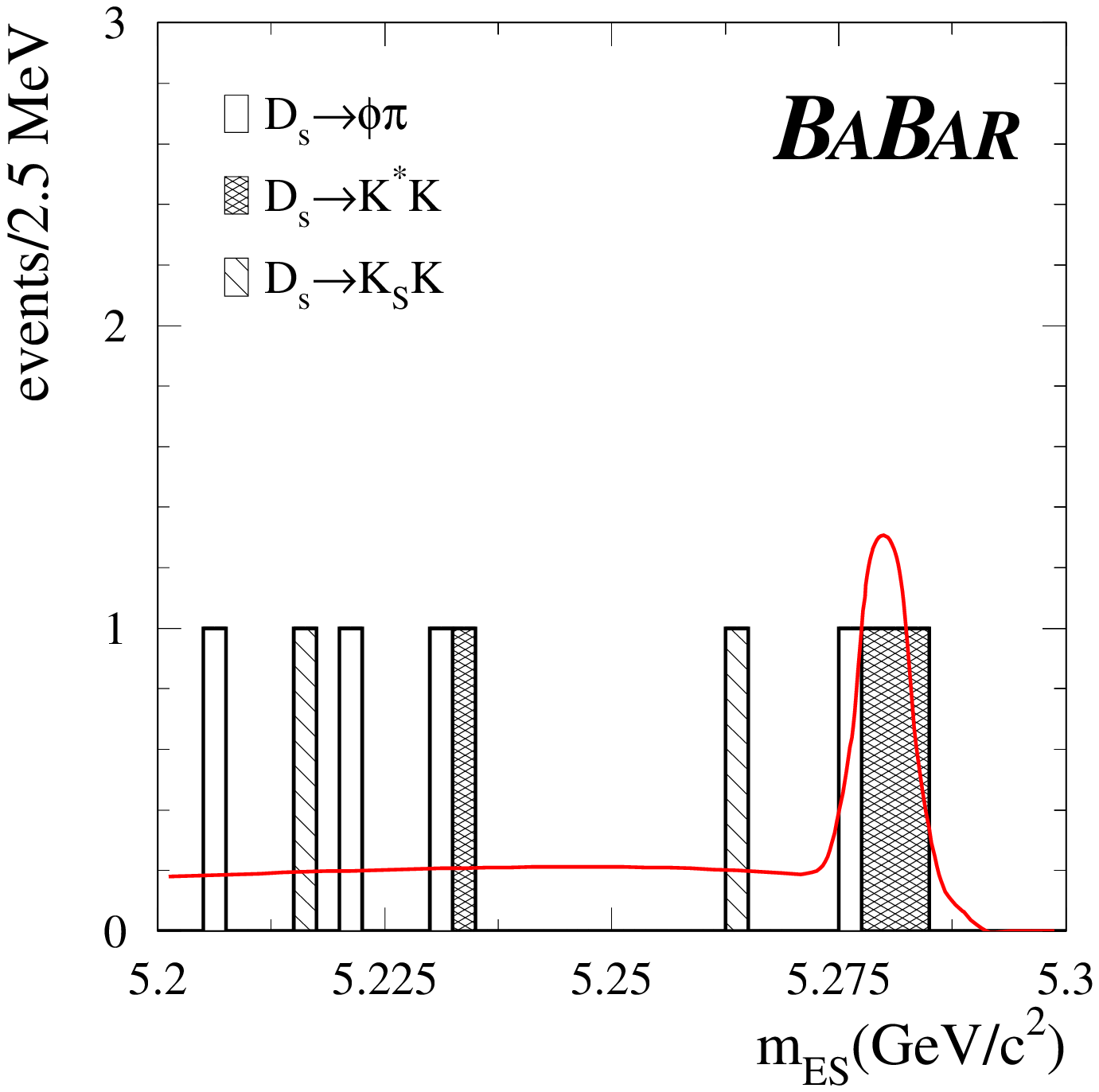}}
    \caption{$m_{\rm ES}$ distribution for $B^0 \to D_s^+ \pi^-$ (left) and $B^0 \to D_s^{*+} \pi^-$ (right)}
\label{Fi:dspi}
\end{center}
\end{figure}

Fig. \ref{Fi:dspi} show the $m_{\rm ES}$ distributions obtained with
56.4 fb$^{-1}$ on-resonance data. The mass distributions are fitted with a Gaussian
function for the signal and a so-called ``ARGUS'' shape function for the background \cite{ref:argus}. 

A signal of $14.9\pm 4.1$ events is found for the $B^0\to D_s^+ \pi^-$ decay mode
with a statistical significance of 3.5 $\sigma$. This corresponds to a branching
fraction of:
\begin{equation}
{\cal B}\left(B^0\to D_s^+ \pi^- \right) = 
\left(3.1 \pm 1.0 {\rm(stat)} \pm 1.0 {\rm(syst)}\right)\times 10^{-5}
\end{equation}

No significant signal is found for the $B^0\to D_s^{*+} \pi^-$ decay mode. An upper
limit at 90 \% of confidence level is derived:
\begin{equation}
{\cal B} \left(B^0\to D_s^{*+} \pi^- \right) < 4.3 \times 10^{-5} 
\end{equation}

\section{Conclusions}

Preliminary results from the {\sc BaBar} experiment have been presented concerning the
branching fractions for the decay modes $B^-\to D_{CP}^0 K^-$, $B^0\to D^{*-} a_1^+$ and
$B^0\to D_s^{(*)+}\pi^-$. These studies show the feasibility of the analyses but 
more statistics are needed to have access to the $\gamma$ angle of the unitarity
triangle.




\end{document}